# Machine learning thermal circuit network model for thermal design optimization of electronic circuit board layout with transient heating chips

Daiki OTAKI[1], Hirofumi NONAKA[2], Noboru YAMADA[1,*]

[1] Department of Mechanical Engineering, Nagaoka University of Technology, 1603-1, Kamitomioka, Nagaoka, Niigata, 940-2133

[2] Department of Information and Management Engineering, Nagaoka University of Technology, 1603-1, Kamitomioka, Nagaoka, Niigata, 940-2133

*Corresponding author: noboru@nagaokaut.ac.jp

**Abstract**

This paper describes a method combining Bayesian optimization (BO) and a lamped-capacitance thermal circuit network model that is effective for speeding up the thermal design optimization of an electronic circuit board layout with transient heating chips. As electronic devices have become smaller and more complex, the importance of thermal design optimization to ensure heat dissipation performance has increased. However, such thermal design optimization is difficult because it is necessary to consider various trade-offs associated with packaging and transient temperature changes of heat-generating components. This study aims to improve the performance of thermal design optimization by artificial intelligence. BO using a Gaussian process was combined with the lamped-capacitance thermal circuit network model, and its performance was verified by case studies. As a result, BO successfully found the ideal circuit board layout as well as particle swarm optimization (PSO) and genetic algorithm (GA) could. The CPU time for BO was 1/5 and 1/4 of that for PSO and GA, respectively. In addition, BO found a non-intuitive optimal solution in approximately 7 minutes from 10 million layout patterns. It was estimated that this was 1/1000 of the CPU time required for analyzing all layout patterns.

## 1. Introduction

In recent years, electronic devices and energy storage systems have become smaller (thinner), and their internal structures have become more complex; therefore, ensuring the heat dissipation performance of these devices has become an issue. Thermal design and temperature control are becoming key elements in improving product performance and reliability[1][2]. For example, temperature suppression methods combining composites, fins, fluids, and phase change materials (PCMs) amongst others, have been studied previously [3][4][5][6][7][8][9][10][11]. Most of the heat generated inside a small electronic device is dissipated through the circuit board (CB) on which the heat-generating components are mounted, and then dissipated to its surroundings. To improve this heat dissipation, i.e., cooling performance, it is necessary to optimize the layout of the heat-generating components mounted on the CB (hereinafter referred to as the CB layout). This optimization process requires consideration of trade-offs due to various constraints in product packaging and a large number of layout patterns, making the search for an optimal solution difficult. Therefore, the development of efficient thermal design methods using artificial intelligence (AI) has been reported [12][13][14][15][16][17][18][19]. Lianlian et al. [12] used ant colony optimization (ACO) to optimize the printed CB layout. Alexandridis et al. [13] applied particle swarm optimization (PSO) to a similar CB layout optimization. Ismail et al. [14] used a genetic algorithm (GA) for a similar CB layout optimization. The reported studies show the feasibility of metaheuristic algorithms to the CB layout optimization although these algorithms have issues such as convergence to a local solution, difficulty to determine appropriate hyperparameters, depending on the optimization problems [20][21][22][23].

Bayesian optimization (BO) using Gaussian processes, known as a type of machine learning, has recently been applied to achieve the optimal design of various devices and systems [24][25][26][27][28][29][30][31]. It has been reported that BO does not easily fall into a local solution [32][33], and its algorithm is provided as a programming library for ease of use. Therefore, BO may be effective for the CB layout optimization problem; however, its effectiveness has not been verified so far, to the best of the authors' knowledge.

In this study, BO is combined with a lamped-capacitance thermal circuit network model and applied to CB layout optimization problems, and its effectiveness is verified by comparing it with other algorithms (PSO and GA). Furthermore, the optimization is carried out based on unsteady state temperature simulations in which time variations of heating power and temperature of the components are considered. In the reported studies on CB layout optimization, the heating power of the heating components is assumed to be constant, and the layout optimization is performed using the temperature under steady-state conditions. However, in actual CBs, the heating power often varies with time. Therefore, the present study unveils the performance potential of BO in more complicated optimization cases than the reported studies.

## 2. Simulation model and problem setting
### 2.1 Thermal circuit network model for CB layout optimization

A lumped-capacitance thermal circuit network model (LTCNM) was used for the transient temperature simulation of a CB. LTCNM is a model based on the analogy between electrical circuits and heat transfer phenomena and is widely used in the design simulation of thermal systems because it allows unsteady heat transfer simulation of systems without fine computational mesh and time-consuming complex fluid simulation [34][35][36][37][38][39][40].

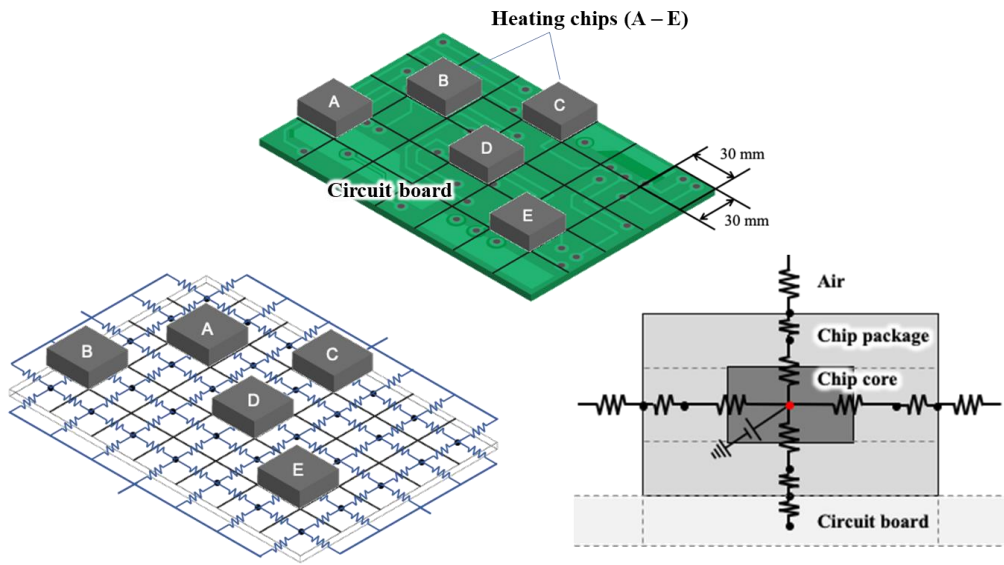

Fig. 1 Lumped-capacitance thermal circuit network model for circuit board with five heating chips

Figure 1 shows the computational CB model to be optimized. There are five heat-generating components (hereafter referred to as heating chips or chips) on the CB. The dimensions, maximum heating power, and integrated heating energy (for $0 \leq t \leq 1800$ s) of the CB and each chip are summarized in Table 1. The thermal circuit network model for the CB consists of 5×7 computational nodes. The lower right figure shows the thermal circuit network model for the heating chip. It consists of a central semiconductor chip core (red-colored 1 node), which is the heat source, and its surrounding resin chip package (6 nodes). The thermophysical properties of the model components are summarized in Table 2. As shown in Fig. 2, it was assumed that the heating power of each chip varied with time. Each chip has different time-varying characteristics, and the total heating power, which is the sum of the heating power of each chip (black line), and peaks at $t = 1047$ s.

Table 1 Specification of circuit board model components

|  | Dimensions [mm] | Max. heating power [W] | Integrated heating energy [Wh] |
|---|---|---|---|
| Board | 210 × 150 × 1 | N/A | N/A |
| Chip A | 30 × 30 × 10 | 4.1 | 1.54 |
| Chip B | 30 × 30 × 10 | 4.0 | 1.25 |
| Chip C | 30 × 30 × 10 | 3.8 | 1.03 |
| Chip D | 30 × 30 × 10 | 3.0 | 0.83 |
| Chip E | 30 × 30 × 10 | 2.0 | 0.50 |

Table 2 Thermophysical properties of circuit board model components

|  | Board | Chip package | Chip core |
|---|---|---|---|
| Materials | Aluminum | Epoxy resin | Silicon |
| Specific heat [J/g·K] | 0.9 | 1.5 | 0.77 |
| Density [g/cm$^3$] | 2.7 | 1.2 | 2.3 |
| Conductivity [W/m·K] | 170 | 0.3 | 156 |

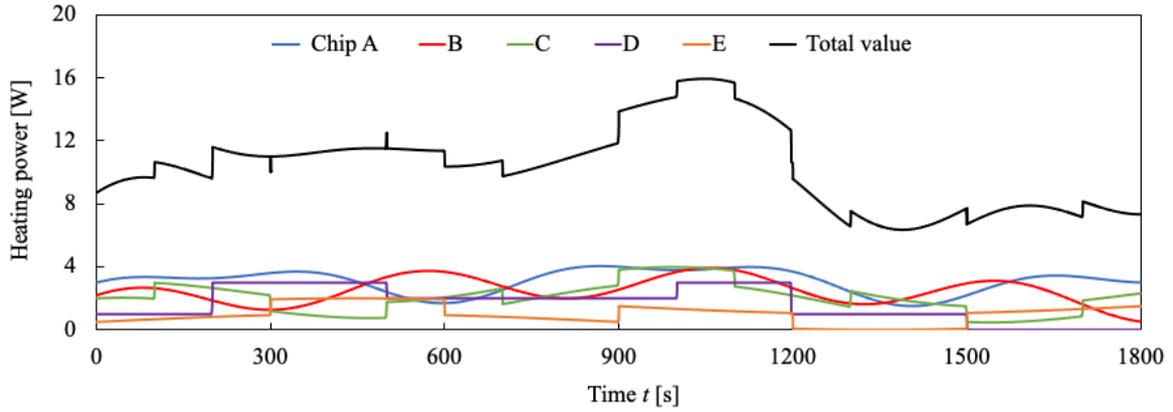

Fig. 2 Time-variations of heating power of five heating chips (A-E) in circuit board model

Because the lumped-capacitance model is used, the heat capacitance is connected to every node in Figure 1, although their circuit symbols are omitted in the figures. It is noted that the circuit symbols of thermal resistance between the nodes and the ambient air are also omitted. In LTCNM, the temperature at the $n$th node is calculated as:

$$\sum_{i=1}^{l} Q_i = \frac{dT_n}{dt} m_n c_n \tag{1}$$

where $Q_i$ [W] is the amount of heat flowing from the neighboring nodes, $l$ is the number of neighboring nodes, $T_n$ [K] is temperature, $t$ [s] is time, $m_n$ [kg] is mass, $c_n$ [J/kg·K] is the specific heat, and the subscript indicates the number of nodes. $Q_i$ is calculated by the following equation:

$$Q_i = \frac{\Delta T_i}{R_i} \tag{2}$$

$\Delta T_i$ [K] is the temperature difference between the *n*th node and the adjacent node, $R_i$ [W/K] is the thermal resistance considering conduction and convection [36]. The initial and boundary conditions are listed in Table 3. This model was implemented in MATLAB/Simulink, and an unsteady heat transfer simulation was performed to obtain the temperature change characteristics of each node. The chip temperature is defined as the average temperature over the nodes within the chip component.

Table 3  Simulation conditions

| | |
|---|---|
| Initial temperature of all components | 25 °C |
| Air temperature | 25 °C (Constant) |
| Boundary conditions | Board-Air: Natural convection<br>Chip-Air: Natural convection<br>Board side and bottom: Adiabatic |
| Heat transfer Coefficient for natural convection | 10 W/m$^2$·K |

Prior to performing the optimization and to confirm the validity of the present LTCNM, three-dimensional (3D) finite element method (FEM) simulations under the same model and conditions were conducted, and the results were compared with the results obtained by LTCNM. The commercial software ANSYS was used for the FEM simulation. FEM simulation is a widely used simulation in the field of heat transfer engineering. The FEM simulation generally provides reliable results; however, it requires a fine 3D computational mesh and a fine time step, resulting in high computational cost. The number of computational meshes used in the FEM model was set to 3630. By contrast, the number of computational nodes in LTCNM is 95. It was confirmed that the time-variation of the volume average temperature of the chip component in both simulations agreed well although the spatial resolution of temperature distribution of LTCNM is lower than that of FEM simulations.

2.2  Optimization problem setting

The target of the optimization is the CB layout, that is, the placement pattern of five transient heating chips A to E. In actual product design, there may be restrictions on chip placement depending on the functions of the devices. To simulate this situation, two restrictions on chip placement are given as i) placeable area for each chip and ii) distance between chips. In restriction i), each chip can only be placed in a mesh defined by a frame of the same color as the color of the chip symbol, as shown in Figure 3. In other words, each chip can only be placed within a specified area. In restriction ii), the distance between the nodes at the center of the chips must be less than or equal to the values shown in Table 4. For example, in Figure 4, the distance between chips A and B is 51.96 mm, which satisfies the restricted value (90 mm).

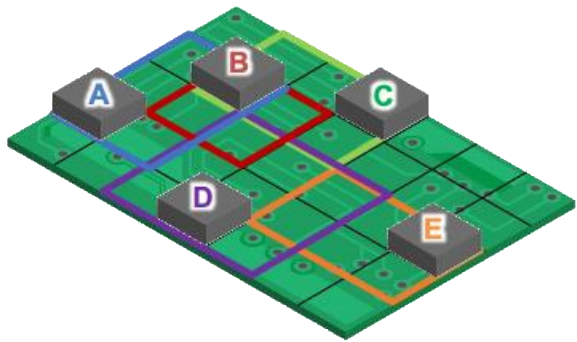

Fig. 3  Restriction i): Colored mesh depicts placeable areas for each chip with the same colored symbols

Table 4  Restriction ii)

| | Max. distance between the chips [mm] |
|---|---|
| Chips A-B | 90 |
| Chips B-C | 90 |
| Chips A-D or D-E | 90 |

The optimization was performed so that the value of the following objective function $f(x)$ was minimized:

$$f(x) = w \max\{T_{\text{mean}}(t)\} + (1 - w) \max\{T_{\text{high}}(t)\} \tag{3}$$

where $x$ indicates one of the CB layouts. $T_{\text{mean}}(t)$ is the mean chip temperature, that is, the average chip temperature over the five chips at time $t$; $T_{\text{high}}(t)$ is the highest chip temperature among the five chips at time $t$; max{ } is the maximum value during the simulation time period ($0 \le t \le 1800$ s); $w$ is the weight coefficient; with the optimization being performed for three cases namely: $w = 1, 0$, and $0.5$. This methodology is called the weighted sum method and is commonly used for multi-objective optimization [41][42]. The objective function for $w = 1, 0$, and $0.5$ is $f(x) = \max\{T_{\text{mean}}(t)\}$, $f(x) = \max\{T_{\text{high}}(t)\}$, and $f(x) = [\max\{T_{\text{mean}}(t)\} + \max\{T_{\text{high}}(t)\}]/2$, respectively.

## 3. Applying Bayesian optimization

The thermal design optimization problem can be formulated as an optimization of the black box continuous functions $f(x)$ as follows:

$$x = \operatorname{argmin}_x f(x) \tag{4}$$

where $x$ is the input variable and $f(x)$ denotes the objective function shown in Equation (3). The de-facto standard model for black-box optimization is the Bayesian optimization (BO) with a Gaussian process. Bayesian optimization (BO) is a popular framework for optimizing the black box function owing to its sample efficiency. The Gaussian process has been widely applied to solve real-world problems such as the prediction of thermal systems because of its ability to capture non-linearity and quantify uncertainty [43][44][45]. Due to such characteristics, a Gaussian process is often selected to model the unknown objective function of BO. In BO, $f(x)$ is a stochastic process, which is assumed to follow a Gaussian process, that is, the following equation (5), where $\mu(x)$ is the mean function (of the objective function at point $x$), $\sigma(x)$ is the covariance function (of the objective function at a point $x$), and $k(x, x')$ is the kernel function.

$$f(x) \sim GP(\mu(x), \sigma(x)) = GP(\mu(x), k(x, x')) \tag{5}$$

In this method, the posterior distribution of $f(x)$ is calculated from the currently observed data based on Equation (5), and the next search point is determined using the acquisition function based on the information of the peripheralized predicted distribution. This process is repeated to find the optimal solution [46][47][48]. The Matérn5/2 kernel [33] was used as well as an expected improvement (EI) [32][33]. The combination of EI and the Matérn5/2 kernel is often used in practical applications [29]. In simulation, the "bayesopt" function of MATLAB library was used. Optimization with PSO and GA was also performed and compared with BO. The "particleswarm" and "ga" functions of the MATLAB library were used for PSO and GA, respectively. The hyperparameters of PSO and GA are shown in Table 5, which were the typical hyperparameter settings for PSO and GA [49][50][51]. The total number of iterations was set to 200 for all algorithms. To determine the true optimal CB layout (henceforth, ideal layout), all 7776 layout patterns were searched and $f(x)$ for them were evaluated in advance. The performance of each algorithm was verified by comparing their optimized layout with the ideal layout. The CPU time consumed by each algorithm was compared. All the simulations were conducted on a Windows workstation with AMD Ryzen9 3950X 3.49 GHz and 16 GB memory. Note that previous studies [12][13][14][15][16][17][18] have not made comparisons with the ideal layout.

Table 5 Hyperparameters of GA and PSO

| | | |
|---|---|---|
| **GA** | | |
| | Population size | 50 |
| | Selection | Stochunif |
| | Mutation rate | 0.01 |
| | Crossover rate | 0.8 |
| **PSO** | | |
| | Swarm size | 50 |
| | Cognitive parameter | 1.49 |
| | Social parameter | 1.49 |

## 4. Optimization results

Tables 6 and 7 show the comparison of optimization results, that is, values of $f(x)$ of the optimized CB layout, and CPU time, for each algorithm including all layout pattern searches. With 200 iterations, three algorithms reached the same layout as the ideal layout regardless of the $f(x)$ case. BO reached the ideal layout in approximately 1/150–1/90 of CPU time for all layout pattern searches, and in approximately 1/5 and 1/4 of the CPU time of PSO and GA, respectively. The swarm size and the population size affect the CPU time of PSO and GA, respectively. In this case, because both sizes are set to 50, PSO and GA must evaluate 50 layout patterns per iteration. By contrast, BO evaluated one layout pattern per iteration. As the transient temperature simulation per layout pattern takes a long CPU time, PSO and GA must take a longer CPU time per iteration than BO. This is the major reason for the increase in CPU time for PSO and GA. The performance of PSO and GA may be improved by careful tuning of hyperparameters, however, such a tuning itself is time consuming. These results imply that BO has a potential to be a time-efficient algorithm for the CB layout optimization problem coupled with the transient temperature simulation.

Table 6 Comparison of optimization results (temperature).

| Objective function (Optimization to minimize $f(x)$) | $f(x)$ of the optimized CB layout [°C] | | | |
|---|---|---|---|---|
| | Searching all layout patterns | BO | PSO | GA |
| $f(x) = \max\{T_{mean}(t)\}$ | 76.88 | ← | ← | ← |
| $f(x) = \max\{T_{high}(t)\}$ | 90.20 | ← | ← | ← |
| $f(x) = [\max\{T_{mean}(t)\} + \max\{T_{high}(t)\}]/2$ | 83.68 | ← | ← | ← |

Table 7  Comparison of optimization results (CPU time).

| Objective function (Optimization to minimize $f(x)$) | CPU time to obtain optimized CB layout [s] | | | |
|---|---|---|---|---|
| | Searching all layout patterns | BO | PSO | GA |
| $f(x) = \max\{T_{mean}(t)\}$ | 9836 | **183** | 866 | 691 |
| $f(x) = \max\{T_{high}(t)\}$ | ↑ | **163** | 738 | 438 |
| $f(x) = [\max\{T_{mean}(t)\} + \max\{T_{high}(t)\}]/2$ | ↑ | **62** | 328 | 495 |

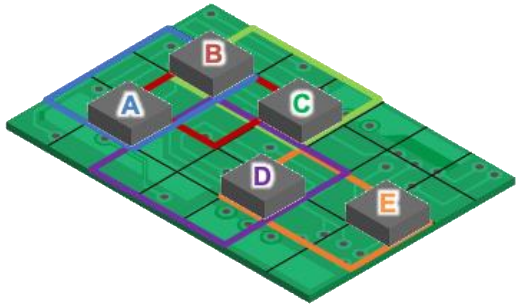
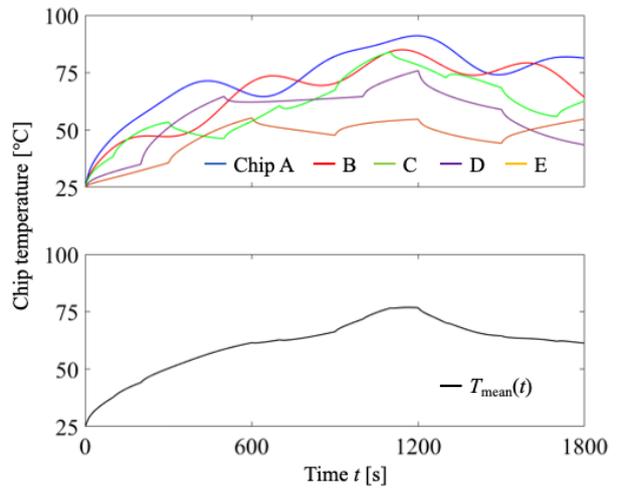

(a)  $f(x) = \max\{T_{mean}(t)\}$

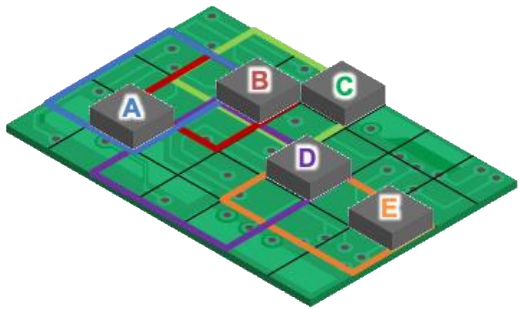
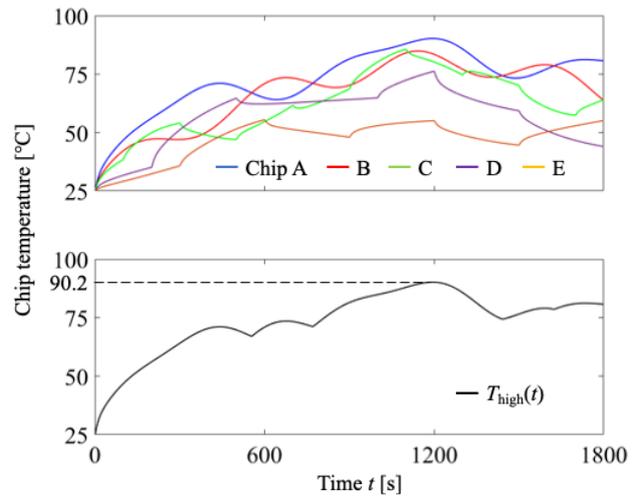

(b)  $f(x) = \max\{T_{high}(t)\}$

Fig. 4  Comparison of optimized component layouts by BO for different objective functions.

Figure 4(a) and (b) show the optimized layouts (which are identical to the ideal layouts) obtained by BO in three $f(x)$ cases, respectively. In the figures, the corresponding time variations of each chip temperature and $T_{mean}(t)$ or $T_{high}(t)$ are also shown. In Fig. 4(a), $T_{mean}(t)$ reaches the maximum value at approximately $t = 1200$ s. This maximum value is the temperature of Chip A, which has the highest total heat generation, as shown in Table 1. In this case, because the chips should be evenly distancing to minimize $f(x) = \max\{T_{mean}(t)\}$, it would be easy for us to predict a similar layout pattern intuitively. By contrast, in Fig. 4(b), with the optimization to minimize $f(x) = \max\{T_{high}(t)\}$, Chips B and C are arranged close to each other, and Chip A is mostly distancing from the other chips, which would not be easy for us to predict intuitively. This indicates that the optimized layouts are reasonable and that the thermal design optimization using BO, that is, a type of machine learning, is effective.

## 5. Applying BO to extended problem settings

In the actual thermal design of CBs, the number of layout patterns can be even greater. To test the performance of BO in such a case, BO was applied to an extended problem setting in which the restrictions i) and ii) described in section 2.2 are removed. In this case, the number of possible layout patterns is approximately 10 million. The optimization for $f(x) = \max\{T_{high}(t)\}$ case was performed. Figure 5 shows the evolution of the value of the objective function with respect to the number of iterations. The values of $f(x)$ at the 20, 200, and 2000 iterations are shown in the graph. Figure 6 shows the corresponding optimized layouts and the time variations of the chip temperature and $T_{high}(t)$ for the optimized layout at 2000 iterations. From Fig. 5, the value of $f(x)$ is updated as the number of iterations increases, reaching 90% of the 2000 iterations' value at 200 iterations. $T_{high}$ at 200 and 2000 iterations were reduced by 2.79°C and 2.9°C, respectively, compared to the $T_{high}$ (90.20°C) under restrictions i) and ii), as shown in Table 6 in Chapter 4. Figure 6 shows that Chips A and B were optimized to maintain distance from the other chips. This is because the heating power of Chips A and B is relatively larger, as shown in Table 1, so they are placed farther distancing from each other to lower $T_{high}$. Comparing the temperature trends in Fig. 6(c) and Fig. 4(b), the temperature of Chip A decreases. The CPU times for 200 and 2000 iterations were 434 s and 16676 s, respectively. Based on the results in Chapter 4, the estimated CPU times for 2000 iterations for PSO and GA are 160,000 s, which is approximately 10 times longer than that of BO. Similarly, the CPU time required for the simulation of all layout patterns search was estimated to be about 140 days, and BO was able to optimize at approximately 1/1000 of the time for all layout pattern searches. From these results, the high speed of BO was confirmed in the extended problem setting. However, it is reported that the computational complexity of BO tends to increase with the number of iterations [48]. This fact was also confirmed by the present result. The CPU time for 2000 iterations was 38 times higher than that for 200 iterations. This characteristic should be considered when performing the optimization by BO.

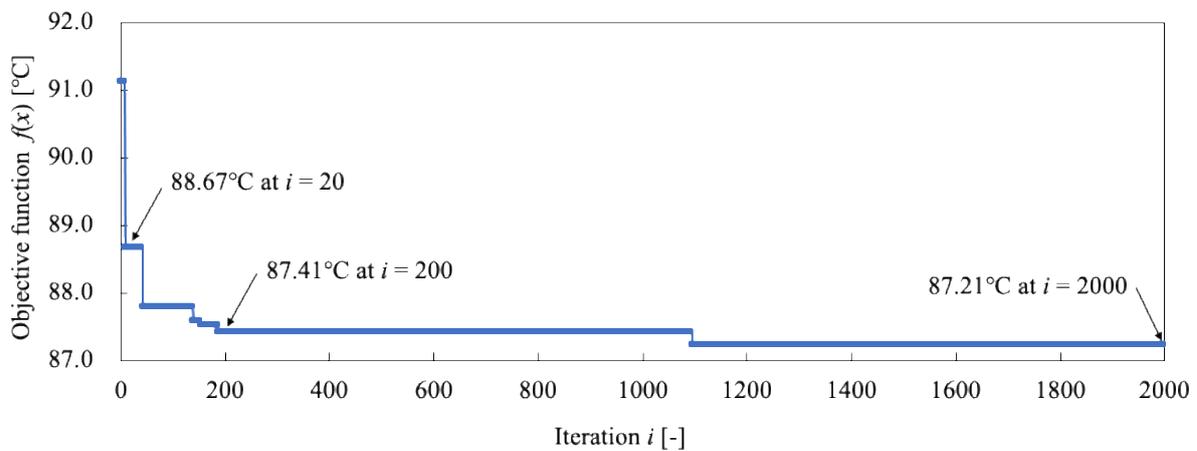

Fig. 5 Optimization progress by BO

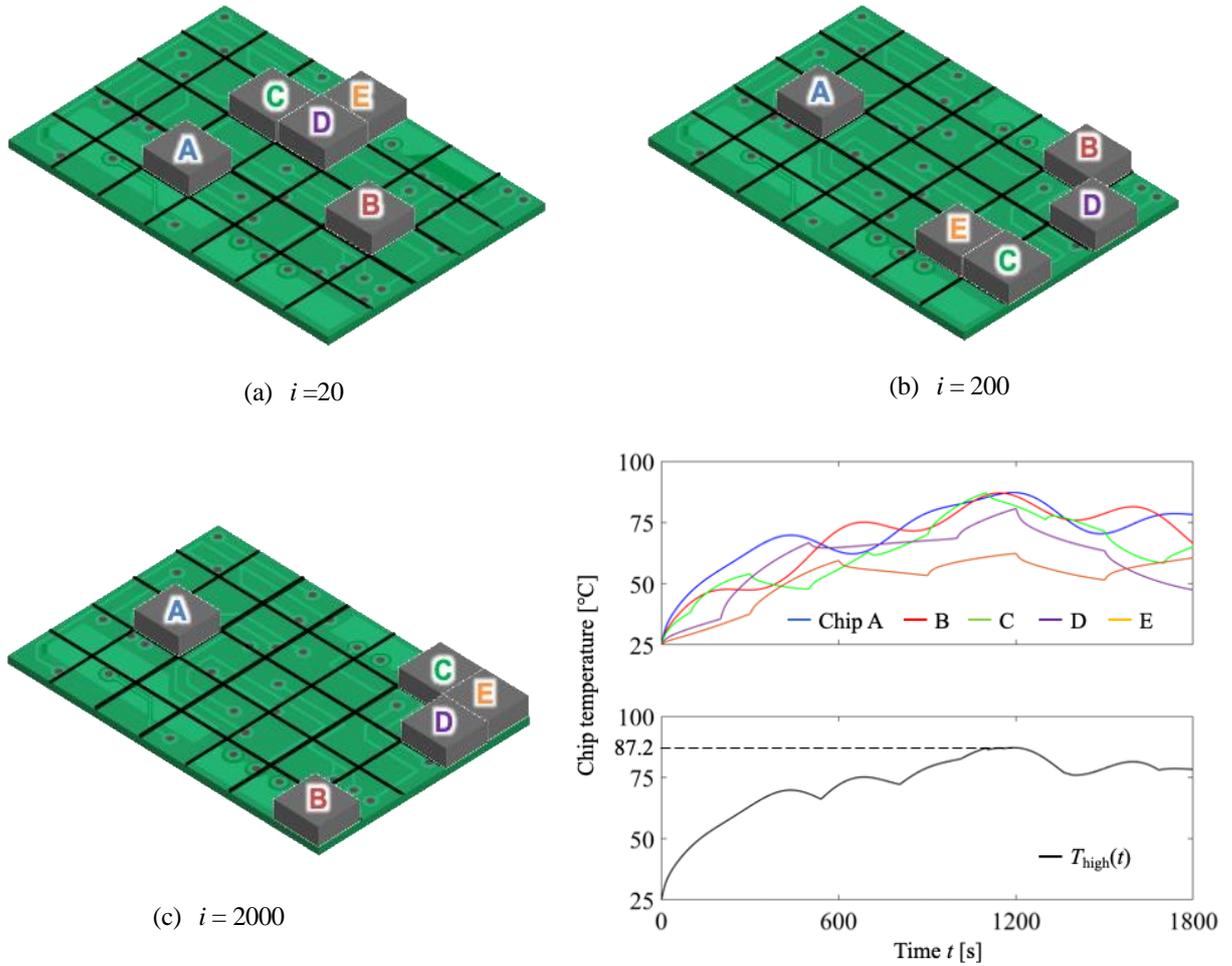

Fig. 6  Optimized component placement by BO w/o restrictions i) and ii).

**6. Conclusion**

   The lamped-capacitance thermal circuit network model combined with BO was applied to the layout optimization of an electronic circuit board with transient heating chips, and its effectiveness was verified by fundamental case studies. To evaluate the value of the objective function of the examined layout, a transient heat transfer simulation was performed per layout taking into account the different temporal variations of the heating power of the heat-generating chips. As a result, BO reached the ideal layout in approximately 1/150–1/90 of CPU time for all layout pattern searches, and in approximately 1/5 and 1/4 of the CPU time of other algorithms (PSO and GA), respectively. Furthermore, BO was applied to the extended problem setting with possible layout patterns of 10 million. BO found a reasonably best layout, which achieved 90% of the objective function value of 2000 iterations, at 200 iterations. It took only approximately 7 minutes. All layout pattern searches would have required 140 days, and BO took only 1/1000 of that time for optimization. In addition, by comparing it with other algorithms (PSO and GA), the effectiveness of the present method (lamped-capacitance thermal circuit network model + BO) was demonstrated for the CB layout optimization problem that requires a transient heat transfer simulation. In future research, it will be necessary to upgrade to a more realistic model that also considers more complicated circuit board structures and conditions.

## Declarations

### Funding
No funding is received for this research.

### Conflicts of interest
The authors declare that they have no conflict of interest.

### Availability of data and material
All data and material is available upon request.

### Code availability
Codes are available upon request.